\documentclass[a4paper,11pt]{article}
\usepackage{pos}
\usepackage{lineno}

\newcommand{\pt}{p_{\text{T}}}
\newcommand{\TeV}{\text{TeV}}
\newcommand{\GeV}{\text{GeV}}
\newcommand{\PbPb}{\text{PbPb}}
\newcommand{\pPb}{\text{pPb}}
\newcommand{\pp}{\text{pp}}

\newcommand{\minv}{m_{\text{inv}}}
\newcommand{\vn}{v_{\text{n}}}
\newcommand{\vTwo}{v_{2}}
\newcommand{\vThree}{v_{3}}

\newcommand{\roots}{\sqrt{s}}
\newcommand{\rootsNN}{\sqrt{s_{_{NN}}}}

\newcommand{\Dzero}{\text{D}^{0}}
\newcommand{\antiDzero}{\overline{\text{D}}^{0}}

\title{Recent results of $\Dzero$ mesons azimuthal anisotropy using the CMS detector}
\ShortTitle{$\Dzero$ mesons azimuthal correlations in CMS}

\manuallySeparateAuthors
\author*[a,b]{Cesar A. Bernardes}
\author{(on behalf of the CMS Collaboration)}

\affiliation[a]{Instituto de F\'{i}sica, Universidade Federal do Rio Grande do Sul (UFRGS),\\
  Av. Bento Gonçalves 9500, Porto Alegre, Brasil}

\affiliation[b]{Instituto de F\'{i}sica Te\'{o}rica, Universidade Estadual Paulista (UNESP),\\
R. Dr. Bento Teobaldo Ferraz 271, S\~{a}o Paulo, Brasil}

\emailAdd{cesar.augusto.bernardes@cern.ch}

\abstract{In a relativistic heavy ion collision, heavy flavor (charm and bottom) quarks are mostly created via hard processes at the early stage of the collisions. We present the latest results of the azimuthal anisotropy coefficients $\vn$ for prompt and nonprompt $\Dzero$ mesons in $\PbPb$, $\pPb$, and $\pp$ collisions from the CMS experiment. The studies are about collectivity phenomena in smaller systems ($\pp$ and $\pPb$ collisions), searches for the effects of very strong electromagnetic fields created in the initial stages of ultrarelativistic $\PbPb$ collisions, and charm quark energy loss in the quark-gluon plasma.}

\FullConference{%
  *** The European Physical Society Conference on High Energy Physics (EPS-HEP2021), ***\\
  *** 26-30 July 2021 ***\\
  *** Online conference, jointly organized by Universität Hamburg and the research center DESY ***
}

\begin{document}
\maketitle

In ultrarelativistic nucleus-nucleus collisions at 
BNL RHIC and CERN LHC, strongly interacting matter with quarks and gluons 
as degrees of freedom (quark-gluon plasma, QGP)
is produced at time scales of ${\sim}1-10~\text{fm}/c$. 
The azimuthal particle correlations observed in such collisions 
are used to study the properties of the QGP. 
A parametrization by a Fourier expansion, with 
Fourier coefficients, $\vn$, gives information 
about the initial collision geometry and its fluctuations. 
The second order ($\vTwo$) coefficient is mostly related to the 
almond-shaped geometry of the overlapping region in the collisions, while 
the third order ($\vThree$) coefficient is associated with fluctuations in the position of the nuclei constituents~\cite{Poskanzer:1998yz}. 
Heavy flavor quarks, charm and bottom, are mostly 
produced via hard parton scattering processes in such collisions, being 
mostly produced right after the collisions. In addition, 
because of their larger masses compared to typical 
temperatures of any stage of the system evolution, 
heavy-flavor quarks are expected to 
experience the full evolution of the collision system until the hadronization phase~\cite{PhysRevC.71.064904}. 

The heavy-flavor quarks are excellent 
probes of the effects from initial stages of the collisions, such as the ones from very strong and transient (${\sim}10^{-1}\text{fm}/c$) electromagnetic fields (EM) hypothesized to be created by spectators and participants of the collisions~\cite{Gursoy:2018yai}. They can also be used as probes for event-by-event flow fluctuations caused by initial conditions at low transverse momentum ($\pt$) and also by variations of particle energy loss at high-$\pt$~\cite{PhysRevC.102.024906}. In addition, heavy-flavor quarks are used in the study of the origin of collectivity phenomena (similar to the ones in nucleus-nucleus collisions) observed in small colliding systems, like proton-proton ($\pp$) and proton-lead ($\pPb$) collisions, with the potential to discriminate contributions to $\vn$ coefficients from initial and final-state effects~\cite{CMS:2020qul}.

In this proceedings, measurements of prompt and nonprompt (from decays of beauty hadrons) $\Dzero$ mesons azimuthal anisotropy are presented. The data from lead-lead ($\PbPb$, at nucleon-nucleon center-of-mass energy of $\rootsNN=5.02~\TeV$), $\pPb$ at $\rootsNN=8.16~\TeV$, and $\pp$ at $\roots=13~\TeV$ collisions at the LHC are collected using the CMS experiment~\cite{CMSdetector}. The events from $\PbPb$ collisions are from minimum bias (MB) triggered data samples, while events from $\pp$ and $\pPb$ collisions are from both MB and high multiplicity (HM) triggers. 
The prompt and nonprompt $\Dzero$ mesons are identified using 
multivariate selections. 
To measure the flow harmonic coefficients, different methods were used depending on the collision system and type of study. For the results in the $\PbPb$ collisions, scalar product and cumulant methods are used to measure $\vn$ from two-and four-particle correlations, while in $\pp$ and $\pPb$ collisions the two-particle correlation method is employed. 
The signal $\vn$ from $\Dzero$ mesons are extracted by fitting the invariant mass ($\minv$) distribution of the $\Dzero$ candidates and $\vn$ 
values as a function of $\minv$.
The main systematic uncertainties are obtained by studying the $\Dzero$ mesons 
reconstruction and identification selection efficiency, the fit modeling of the $\minv$ and $\vn$, the contamination of nonprompt $\Dzero$ particles in the prompt $\Dzero$ mesons samples, and contributions from hadronic jets for the measurements in small colliding systems. For details about the procedures described above, see Refs.~\cite{CMS:2020bnz, CMS-PAS-HIN-20-001, CMS:2020qul}.

\begin{figure}[h!!!!!!]
    \centering
    \includegraphics[width=0.92\linewidth]{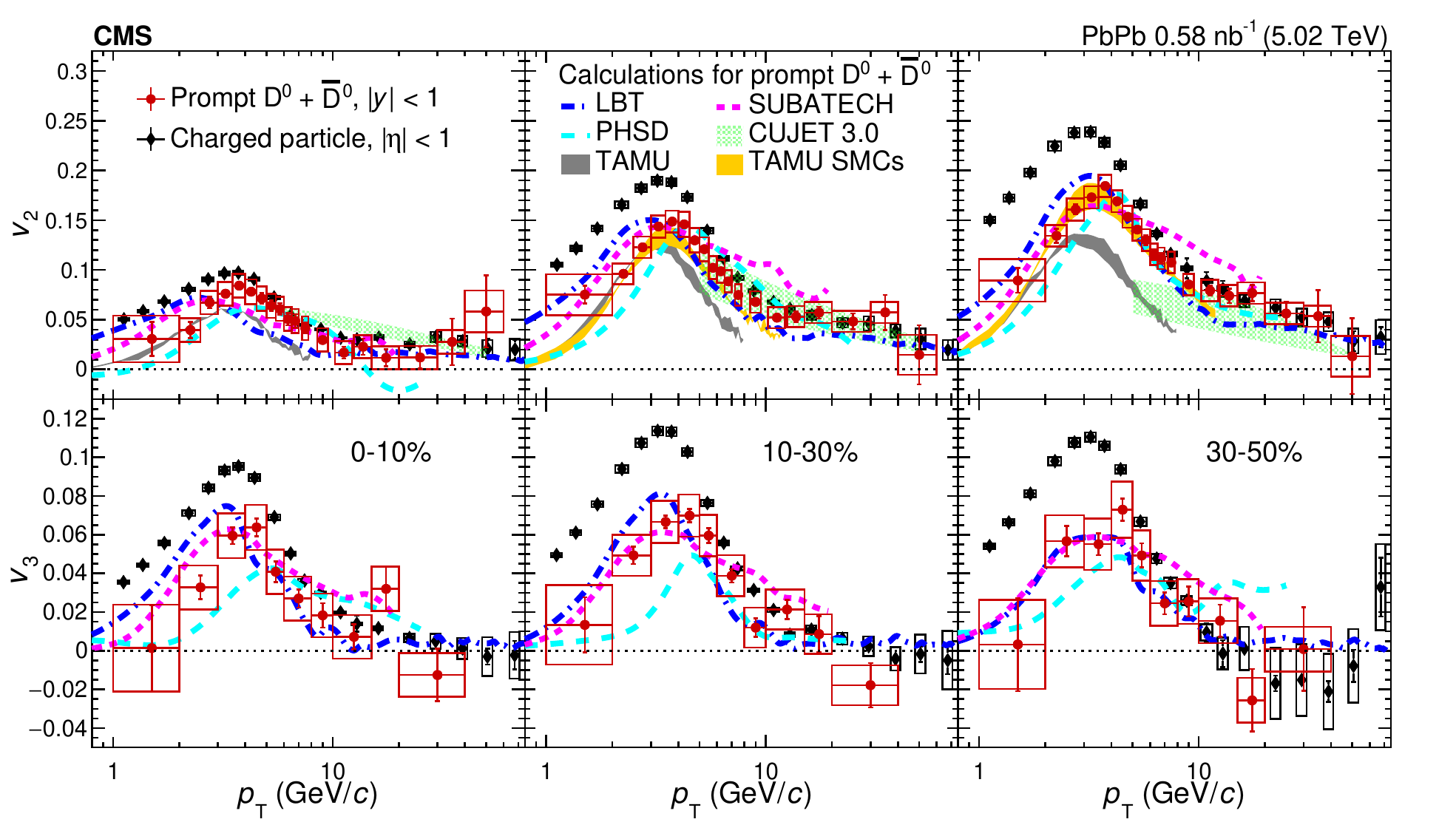}
    \caption{Flow coefficients $\vTwo$ (upper) and $\vThree$ (lower) of prompt $\Dzero$ mesons (with rapidity $|y|<1.0$) and charged particles (with pseudorapidity $|\eta|<1.0$) in $\PbPb$ collisions at $\rootsNN=5.02~\TeV$. The vertical bars are statistical uncertainties, while open boxes represent systematic uncertainties. Several theoretical calculations of $\vn$ of prompt $\Dzero$ mesons are shown for comparison~\cite{CMS:2020bnz}.}
    \label{fig:fig01}
\end{figure}

The $\vTwo$ and $\vThree$ results in Fig~\ref{fig:fig01} extend previously published data from CMS~\cite{CMS:2017vhp}, by enlarging the $\pt$ range up to ${\sim}60.0~\GeV/c$ and by providing finer $\pt$ bins with much smaller statistical uncertainties. Similar trends as in charged particles is observed for $\Dzero$ mesons, 
showing that heavy-flavor quarks flow together with lighter quarks. 
The theoretical calculations also provide good qualitative description of the data, in special, the TAMU SMCs model~\cite{He:2019vgs} shows considerable improvements with respect to TAMU model~\cite{He:2014cla} in the $\pt$ range $3-10~\GeV/c$, by considering space-momentum correlations (SMCs) between charm quarks and high-flow quarks in the QGP.

\begin{figure}[h!!!!!!]
    \centering
    \includegraphics[width=0.45\linewidth]{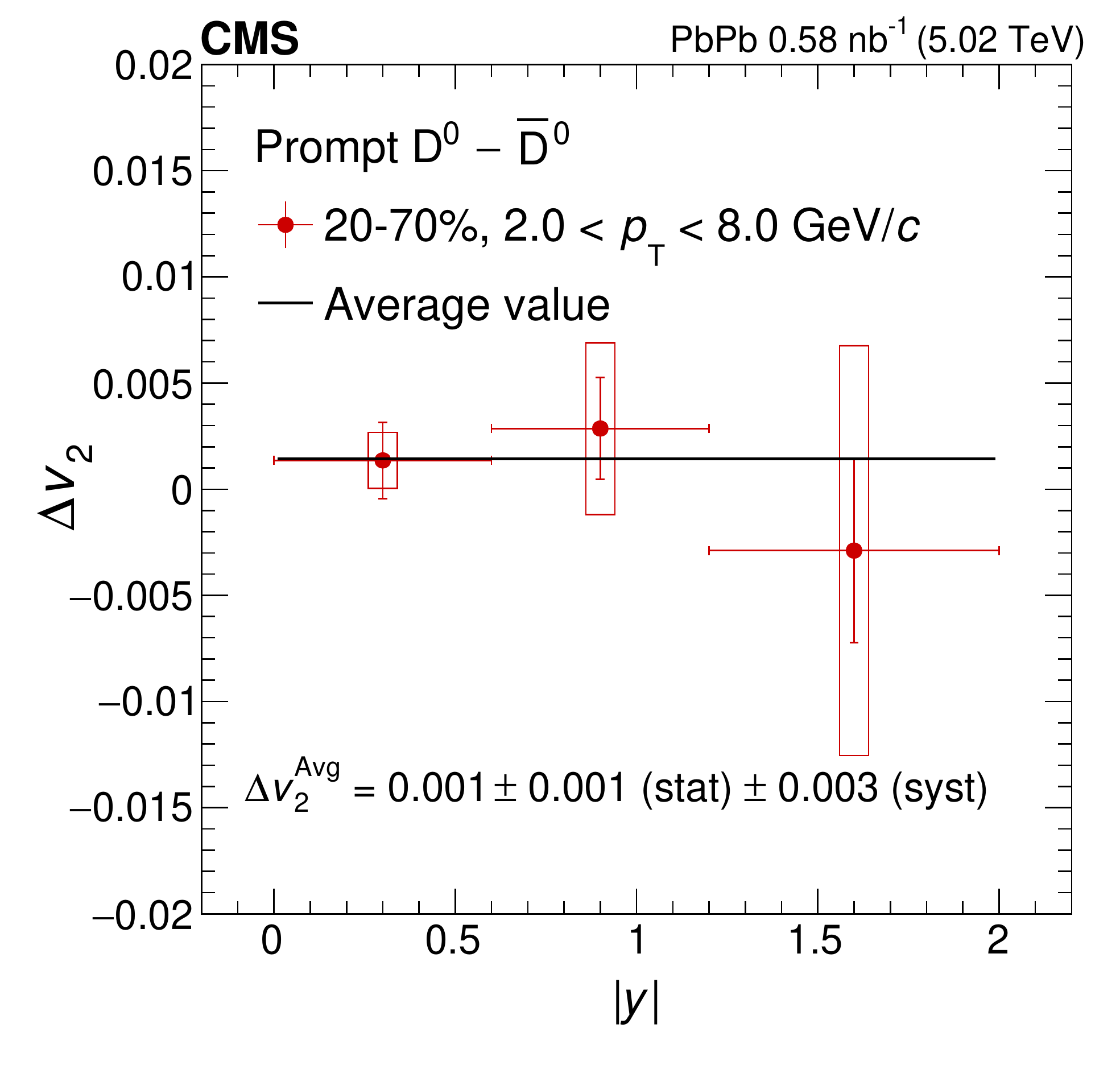}
    \caption{Prompt $\Dzero$ mesons $\Delta\vTwo$ (between $\Dzero$ and $\antiDzero$ particles) as a function of rapidity in $\PbPb$ collisions at $\rootsNN=5.02~\TeV$. The vertical bars are statistical uncertainties, while open boxes represent systematic uncertainties~\cite{CMS:2020bnz}.}
    \label{fig:fig02}
\end{figure}

In ultrarelativistic nucleus-nucleus collisions, very strong and 
transient (${\sim}10^{-1}\text{fm}/c$) EM fields are hypothesized to be 
induced by the collision participants and spectators. There are predictions indicating that such strong EM fields can produce a difference in the $\vn$ coefficients for positively and negatively charged particles. In particular, the Coulomb field created by collision participants is expected to create a difference in $\vTwo$~\cite{Gursoy:2018yai}. 
To search for such effects, the difference $\Delta\vTwo$ 
between the $\vTwo$ values of $\Dzero$ and $\antiDzero$ mesons is measured, as shown in Fig.~\ref{fig:fig02}. The extracted average in the full rapidity region is $\Delta\vTwo^{\text{Avg}}=0.001\pm0.001\text{(stat)}\pm0.003\text{(syst)}$, which is compatible with zero. 
Currently there is no prediction for such effects in charm quarks $\vn$ coefficients, but they are expected to be larger than for lighter quarks (predicted to be of the order of ${\sim}0.001$ at the LHC energies~\cite{Gursoy:2018yai}). This is because heavy-flavor quarks are in general produced much earlier in the collision than light quarks, when the magnitude of the EM fields is larger~\cite{Chatterjee:2017ahy}. Therefore, such small value of $\Delta\vTwo$ can pose constraints on possible EM effects on charm quarks.

\begin{figure}[h!!!!!!]
    \centering
    \includegraphics[width=0.48\linewidth]{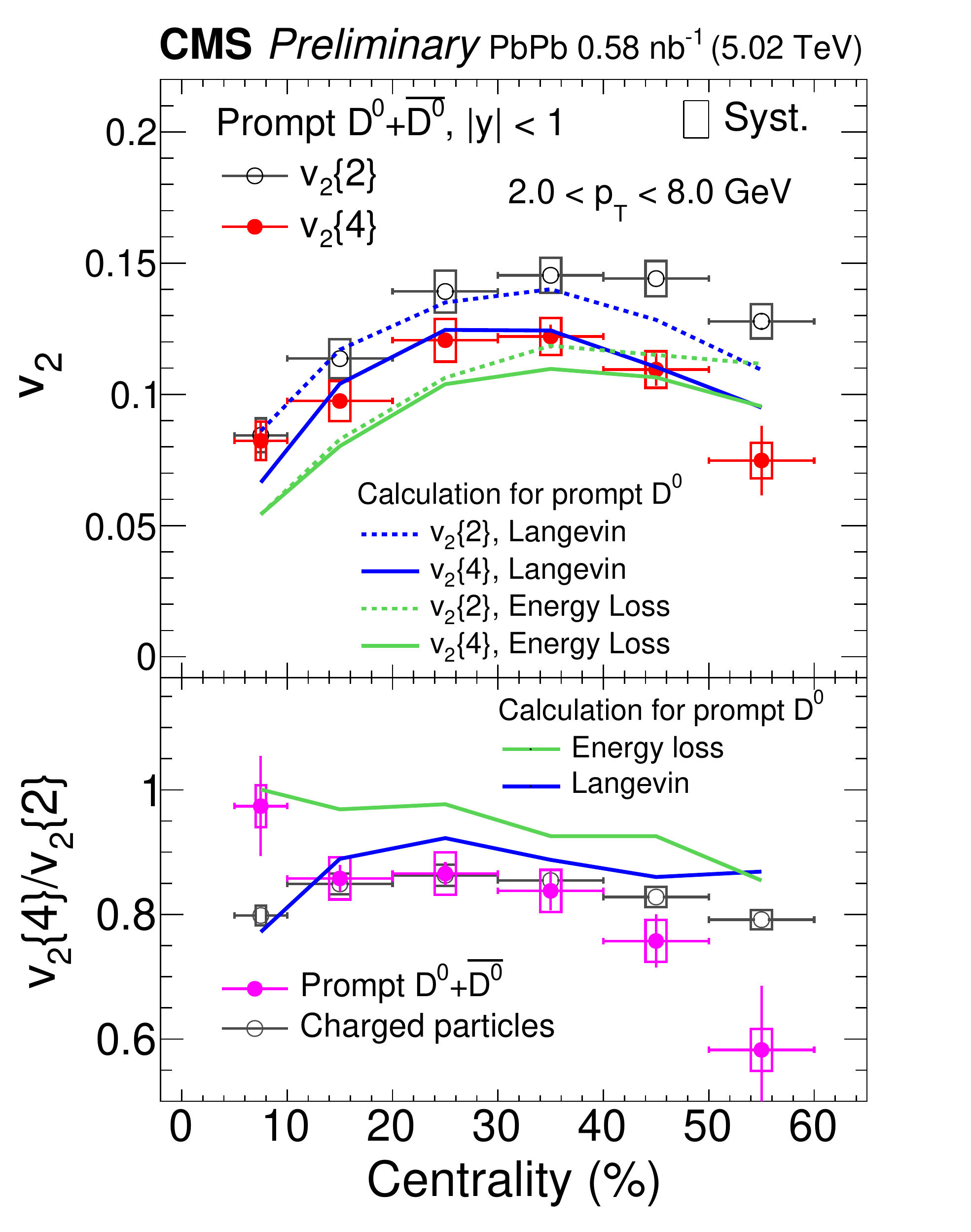}
    \caption{Upper panel: prompt $\Dzero$ mesons $\vTwo\{2\}$ and $\vTwo\{4\}$ as function of centrality in $\PbPb$ collisions at $\rootsNN=5.02~\TeV$. The blue solid and dashed lines denote DABMod model calculations with Langevin dynamics for $\vTwo\{4\}$ and $\vTwo\{2\}$, respectively. The green solid and dashed lines denote DABMod model calculation including radiative energy loss. Lower panel: ratio of $\vTwo\{4\}/\vTwo\{2\}$ for $\Dzero$ mesons ($|y|<1.0$) and charged particles ($|\eta|<1.0$). The vertical bars are statistical uncertainties, while open boxes represent systematic uncertainties~\cite{CMS-PAS-HIN-20-001}.}
    \label{fig:fig04}
\end{figure}

In the Ref.~\cite{CMS-PAS-HIN-20-001}, the $\vTwo$ coefficient of prompt $\Dzero$ mesons is measured for the first time using a four particle cumulant technique, $\vTwo\{4\}$. These measurements allow to access the magnitude of the event-by-event fluctuations of the flow harmonics from heavy-flavor quarks. In particular, the ratio $\vTwo\{4\}/\vTwo\{2\}$ is predicted to be sensitive to energy-loss fluctuations for high-$\pt$ heavy-flavor quarks~\cite{PhysRevC.102.024906}. Figure~\ref{fig:fig04} shows results of $\vTwo\{4\}$ and $\vTwo\{2\}$ (from scalar product method) as a function of centrality from $5$ to $60\%$ and $2<\pt<8~\GeV/c$. The $\vTwo\{4\}$ shows similar trends as $\vTwo\{2\}$, but with smaller magnitude. In the lower panel, the ratio $\vTwo\{4\}/\vTwo\{2\}$ shows a hint of different trends in peripheral events, which could indicate that fluctuations in hard processes become visible for charm mesons, but the uncertainties are large for a firm conclusion. The measurements are compared with DABMod model~\cite{PhysRevC.102.024906}, showing a better description from Langevin dynamics calculations for $2<\pt<8~\GeV/c$. Measurements at higher $\pt$ are also presented in Ref.~\cite{CMS-PAS-HIN-20-001}, but are not shown in this proceedings.

\begin{figure}[h!!!!!!]
    \centering
    \includegraphics[width=0.50\linewidth]{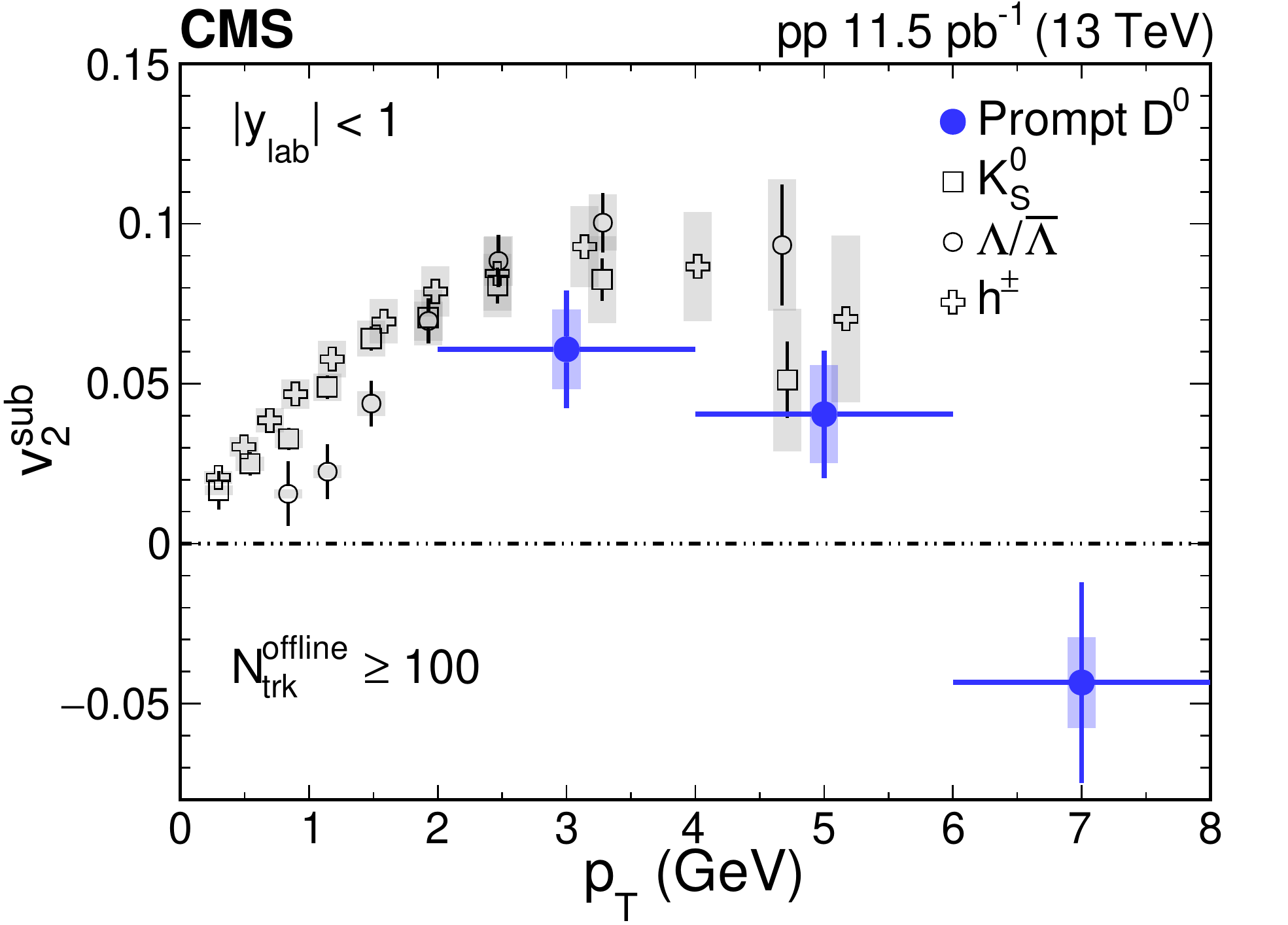}
    \includegraphics[width=0.47\linewidth]{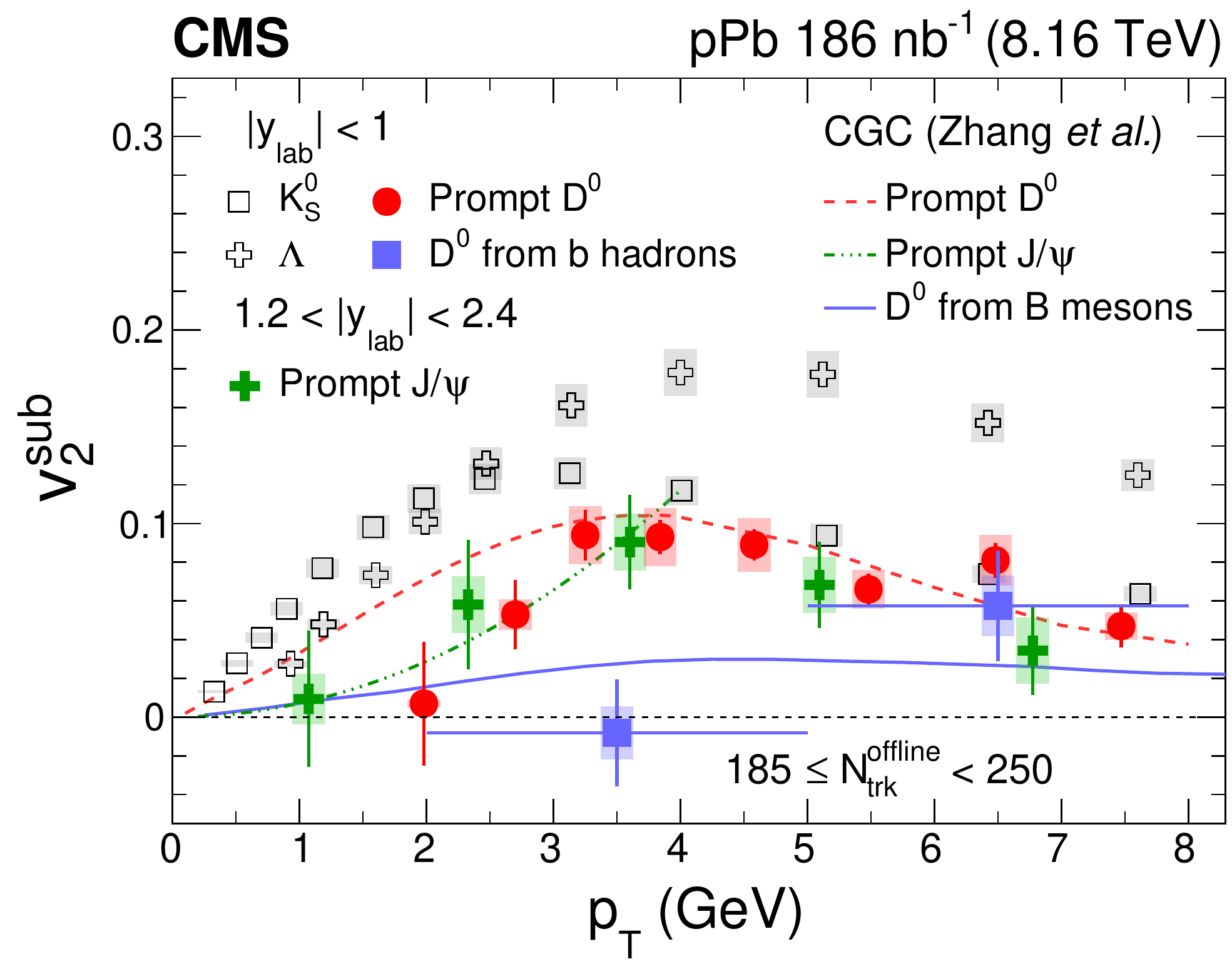}
    \caption{Left: Measurements of $\vTwo^{\text{sub}}$ for prompt $\Dzero$ mesons as a function of $\pt$ in $\pp$ collisions at $\roots=13~\TeV$. Published results for charged particles, $\text{K}_{\text{s}}^0$ mesons, $\Lambda$ baryons are also shown for comparison. Right: Measurements of $\vTwo^{\text{sub}}$ for prompt and nonprompt $\Dzero$ mesons, $\text{K}_{\text{s}}^0$ mesons, $\Lambda$ baryons, and prompt $J/\Psi$ mesons in $\pPb$ collisions at $\rootsNN=8.16~\TeV$. The lines show theoretical calculations within the CGC framework. The vertical bars are statistical uncertainties, while shaded areas represent systematic uncertainties.~\cite{CMS:2020qul}}
    \label{fig:fig05}
\end{figure}

Similarly to nucleus-nucleus collisions, long-range collective azimuthal correlations are observed in small colliding systems with high final-state particle multiplicity ($N_{\text{trk}}^{\text{offline}}$). Figure~\ref{fig:fig05} (left) shows results for the $\vTwo^{\text{sub}}$ (corrected for residual jet correlations~\cite{CMS:2020qul}) of prompt $\Dzero$ mesons in $\pp$ collisions at $\roots=13~\TeV$ as a function of $\pt$. A positive value of the $\vTwo$ signal ($0.061\pm0.018\text{(stat)}\pm0.013\text{(syst)}$) is observed over a $\pt$ range of ${\sim}2-4~\GeV/c$, going to zero at higher $\pt$. This indicates that charm quarks also present such collectivity signatures in $\pp$ collisions. In Ref.~\cite{CMS:2020qul} results of $\vTwo^{\text{sub}}$ as a function of $N_{\text{trk}}^{\text{offline}}$ in $\pp$ and $\pPb$ are also presented in order to investigate possible system size dependence of collectivity for charm quarks.

Figure~\ref{fig:fig05} (right) shows results for nonprompt $\Dzero$ mesons as a function of $\pt$ in $\pPb$ collisions at $\rootsNN=8.16~\TeV$. For $\pt\sim2-5~\GeV/c$, nonprompt $\Dzero$ mesons show a considerably smaller value of $\vTwo$ as compared to prompt $\Dzero$ particles, with a significance of $2.7$ standard deviations. Comparison with calculations from the color glass condensate (CGC) framework~\cite{Zhang:2019dth,Zhang:2020ayy} are also presented, where sizeable $\vTwo$ signals are generated by correlations between partons in the initial stages of the collisions, showing a good qualitative description of the results for $J/\Psi$ mesons and prompt and nonprompt $\Dzero$ mesons.

In summary, recent results on the prompt and nonprompt $\Dzero$ mesons azimuthal anisotropy in $\pp$, $\pPb$, and $\PbPb$ collisions using the CMS experiment are presented. Collectivity effects observed in small systems ($\pp$ and $\pPb$ collisions) are investigated by measuring the elliptic flow coefficient ($\vTwo$) as a function of the $\Dzero$ mesons transverse momentum and flow fluctuation effects in $\PbPb$ collisions are studied by comparing $\vTwo$ measured using two-and four-particle cumulant techniques. In addition, a search for hypothesized electromagnetic fields created at the initial stages of $\PbPb$ collisions is performed by measuring the difference in $\vTwo$ between $\Dzero$ and $\antiDzero$ mesons as a function of rapidity.

\acknowledgments{This material is based upon work supported by the S\~{a}o Paulo Research Foundation (FAPESP) under Grants No. 2013/01907-0 and 2018/01398-1 and by CAPES PRINT Grant No. 88887.468124/2019–00.}

\end{document}